\documentclass[prb,twocolumn,showpacs]{revtex4}
\usepackage{graphicx}
\usepackage{amsmath}
\usepackage{amssymb}
\usepackage{dcolumn}
\usepackage{float}
\usepackage{bm}
\usepackage{lipsum}
\usepackage{blindtext}
\usepackage[breaklinks=true,colorlinks,citecolor=blue,linkcolor=blue,urlcolor=blue]{hyperref}

\DeclareMathAlphabet{\bi}{OML}{cmm}{b}{it}

\def\be{\begin{equation}}
\def\ee{\end{equation}}
\def\bearr{\begin{eqnarray}}
\def\eearr{\end{eqnarray}}
\def\la{\langle}
\def\ra{\rangle}

\begin{document}
\title{Hot electron cooling in Dirac semimetal Cd$_3$As$_2$ due to polar optical phonons}
\bigskip

\author{Shrishail S. Kubakaddi}
\email{sskubakaddi@gmail.com}

\author{Tutul Biswas}
\email{tbtutulm53@gmail.com}
\normalsize
\affiliation
{
$\color{blue}{^\ast}$Department of Physics, K. L. E. Technological University, Hubballi-580 031, Karnataka, India\\
$\color{blue}{^\dagger}$Department of Physics, University of North Bengal, Raja Rammohunpur-734013, India}

\date{\today}

\begin{abstract}
A theory of hot electron cooling power due to polar optical phonons $P_{\rm op}$ is developed in
three-dimensional Dirac semimetal($3$DDS) Cd$_3$As$_2$ taking account of hot phonon effect. Hot phonon
distribution $N_q$ and $P_{\rm op}$ are investigated as a function of electron temperature $T_e$,
electron density $n_e$, and phonon relaxation time $\tau_p$. It is found that $P_{\rm op}$ increases
rapidly (slowly) with $T_e$ at lower (higher) temperature regime.
Whereas, $P_{\rm op}$ is weakly deceasing with increasing $n_e$. The results are compared with those for
three-dimensional electron gas ($3$DEG) in Cd$_3$As$_2$ semiconductor. Hot phonon effect is found
to reduce $P_{\rm op}$ considerably and it is stronger in 3DDS Cd$_3$As$_2$ than in Cd$_3$As$_2$ semiconductor.
$P_{\rm op}$ is also compared with the hot electron cooling power due to acoustic phonons $P_{\rm ac}$.
We find that a crossover takes place from $P_{\rm ac}$ dominated cooling at low $T_e$ to $P_{\rm op}$ dominated
cooling at higher $T_e$. The temperature at which this crossover occurs shifts towards higher values
with the increase of $n_e$. Also, hot electron energy relaxation time $\tau_e$ is discussed and estimated.

\end{abstract}

\pacs{72.10.-d, 73.63.-b 73.20.-r}


\maketitle

\section{Introduction}
Recently, theoretically predicted\cite{Wang_1, Wang_2} and by now experimentally realized and verified three-dimensional
Dirac semimetals (3DDS) have become the rapidly growing field of research
interest\cite{Boris, Liu, Neupan, Jeon, L_He, zk_Liu, T_Liang, YZhao, Weber, H_Weng, Wang_3, W_Lu, Zhu, cp_Weber, C_Zhu}.
These 3DDS, the three-dimensional (3D) analogue of graphene, have gapless band feature with linear
band dispersion and vanishing effective mass in their low energy states. The cadmium arsenide ( Cd$_3$As$_2$),
a potential representative of 3DDS, has drawn more attention as it is robust and chemically stable
compound in air with ultrahigh mobility\cite{Liu, Neupan, T_Liang, YZhao}. The linear band picture of 3D Dirac fermions
has lead 3DDS to exhibit many unusual transport phenomena such as strong quantum oscillations\cite{L_He, YZhao},
ultrahigh mobility\cite{T_Liang, YZhao, J_Feng} and giant magneto resistance\cite{T_Liang, J_Feng, A_Narayan, H_Li}.
Besides, several promising applications 
of bulk Dirac fermions in Cd$_3$As$_2$ in photonic devices such as ultrafast broadband infrared 
photodetectors\cite{Wang_3} and ultrafast optical switching mechanism for the mid-infrared\cite{C_Zhu}
have been realized and demonstrated. Because of the inherent zero energy gap and linear band dispersion,
3DDS can absorb photons in the entire infrared region. These $3$DDS have advantage over
two-dimensional ($2$D) Dirac semimetals like monolayer graphene because bulk nature of $3$DDS enhances
the efficiency of photon absorption.

The experimentally reported low temperature ($\sim 5$ K) high mobilities
$\sim 9\times 10^6$ cm$^2$/V-s \cite{Neupan, Jeon, T_Liang, YZhao} and up to $4.60\times 10^7$ cm$^2$/V-s\cite{YZhao}
are higher than  that in suspended graphene. The measurements
of resistivity $\rho$ vs temperature $T$, show $\rho\sim T$ down to low $T$, which is inferred to be due to umklapp
processes and electron-optical phonon  scattering\cite{YZhao}. High quality $3$DDS Cd$_3$As$_2$ microbelts\cite{ZG_Chen}
and nanobelts\cite{E_Zhang} with room temperature electron mobility $\sim 2\times 10^4$ cm$^2$/V-s have been fabricated.
In nanobelt\cite{E_Zhang}, the Hall mobility $\mu_H$ follows the typical relation $\mu_H \sim T^{-\gamma}$ with $\gamma$=$0.5$,
in the range $20$-$200$ K, which is attributed to the enhanced electron-phonon (el-ph) scattering.

Theoretically, electronic transport properties of $3$D Weyl and Dirac semimetals
are studied using the semi-classical Boltzmann transport equation\cite{Lundgren, SD_Sarma}. 
Considering the electron momentum relaxation processes due to scattering by
disorder (short-range and long-range) and acoustic phonons, the latter is shown
to dominate electrical conductivity at higher temperature\cite{SD_Sarma}. However, 
the quantitative comparison between the existing experimental results and theoretical calculations is still lacking. 

In order to find the applications of $3$DDS in devices operating in the high field region,
it is important to investigate the steady state energy relaxation of the hot carriers, in
these systems, by emission of phonons as the only cooling channel. In high electric field 
electrons gain energy and establish their `hot electron temperature $T_e$' which is greater
than the lattice temperature $T$. In the steady state these hot electrons transfer their
energy to lattice by emission of acoustic (optical) phonons at relatively low (high) temperature. 
The electron heating affects the device operation significantly, in the high field region, 
as it governs the thermal dissipation and heat management. To enhance the device efficiency,
it is important to reduce the hot electron power loss. 

The hot electron energy relaxation by emission of acoustic and optical phonons has been
extensively investigated theoretically and experimentally in
conventional $3$D electron gas ($3$DEG) in bulk semiconductors\cite{Conwell, Seeger, InSb, GaAs,SD_Sarma2, Prabhu, Ridley_1},
$2$D electron gas (2DEG) in semiconductor
heterostructures\cite{Ridley_1, Price}, monolayer graphene\cite{SSK} and bilayer graphene\cite{SSK_B}. 
Recently, hot electron cooling is theoretically studied in monolayer MoS$_2$\cite{Kaasbjrg} and quasi-2DEG in gapped 
Cd$_3$As$_2$ film\cite{Cd_Film}.
There exist theoretical studies of hot $3$D Dirac fermion cooling  power due to 
electron-acoustic phonon interaction $P_{\rm ac}$ in Cd$_3$As$_2$\cite{Lundgren_2, KS_Bhargv}. The  deformation potential
coupling constant $D$ ($\sim 10$-$30$ eV)\cite{Jay_Gerin} determines the strength of electron-acoustic phonon
scattering. In the  Bloch-Gr\"{u}neisen (BG) regime the power laws of $P_{\rm ac}$ dependence on electron
temperature $T_e$ and electron density $n_e$ are predicted\cite{KS_Bhargv}.

Experimentally, the phonon mediated hot electron cooling of photoexcited carriers has been investigated in Cd$_3$As$_2$ from
pump-probe measurements\cite{Weber, W_Lu, Zhu, cp_Weber}. The cooling process of photoexcited carriers is
shown to be through emission of acoustic and optical phonons\cite{Weber, W_Lu}, with relatively
low optical phonon energies $\sim 25$ meV\cite{W_Lu, Weszka}. The hot electron cooling,
apart from relating el-ph scattering to the high field transport properties,
it also plays significant role in designing high speed electronic and photonic
devices of Cd$_3$As$_2$. Thus, el-ph interaction is a key issue and central to the
understanding of devices based on Cd$_3$As$_2$.

It is important to notice that, in $3$DDS Cd$_3$As$_2$, although there is strong experimental evidence
of photoexcited hot carrier energy relaxation by optical phonon emission\cite{Weber, W_Lu, Zhu}, the steady state
hot electron cooling by emission of optical phonons has not been addressed both theoretically and
experimentally. In the present work, we theoretically investigate the hot electron cooling power
in $3$DDS Cd$_3$As$_2$ by emission of optical phonons $P_{\rm op}$ including the hot phonon effect. Numerical
results are obtained as a function of electron temperature, electron density and phonon
relaxation time. These results are compared with $P_{\rm op}$ in bulk Cd$_3$As$_2$ semiconductor and
with $P_{\rm ac}$, in $3$DDS Cd$_3$As$_2$. 
This study is expected to provide thermal link between electrons and phonons in $3$DDS Cd$_3$As$_2$ 
for its application in high speed/field devices.

The structure of the paper is shaped in the following way. In section II we provide 
all the theoretical ingredients including hot phonon effect, cooling power due to 
optical and acoustic phonons in $3$DDS Cd$_3$As$_2$, and cooling power
for $3$DEG in bulk Cd$_3$As$_2$ semiconductor. The obtained results are 
discussed in section III. Finally a summary of the present work is given in section IV.

\section{Theory}
In this section we develop a theory for the cooling power of hot electrons in $3$DDS mediated 
by polar optical phonons. For comparison purpose we shall also provide the 
results for cooling power due to acoustic phonon in $3$DDS as well as that in bulk Cd$_3$As$_2$ semiconductor
due to polar optical phonon. Let us start with mentioning the basic properties of the physical 
system chosen.

\subsection{Preliminary informations}
We consider the Dirac fermion gas in a $3$DDS Cd$_3$As$_2$ in which
the low energy excitations are described by 
the Dirac like linear dispersion $E_{\bf k}=s\hbar v_F \vert {\bf k}\vert$ in the long wavelength continuum limit.
Here, $v_F$ is the Fermi velocity, ${\bf k}$
is the $3$D electronic wave vector, and the band index $s$ takes the value $+1(-1)$ for 
conduction(valence) band. The corresponding eigenstate is given by
$\psi_{\bf k}^s=(1/\sqrt{2V})e^{i{\bf k}\cdot {\bf r}}\chi^s$, where
$V$ is the volume of the system,
$\chi^+=[\cos(\theta/2)~~ \sin(\theta/2)e^{i\phi}]^T$ and 
$\chi^-=[\sin(\theta/2)~~ -\cos(\theta/2)e^{i\phi}]^T$
with $\theta$ and $\phi$ are the polar and azimuthal angle in
three dimensional ${\bf k}$-space, respectively. The corresponding 
density of states is given by $D(E_{\bf k})={\rm g} E_{\bf k}^2/(2\pi^2\hbar^3 v_F^3)$, where 
${\rm g}={\rm g_s}{\rm g_v}$ with ${\rm g_s}({\rm g_v})$ is the spin(valley) degeneracy.

\subsection{Hot electron cooling power in $3$DDS}
In order to formulate a theory for the hot electron cooling power in 3DDS,
we work in the ``hot electron temperature model'' in which the electron gas is
assumed to be in equilibrium with itself at an elevated temperature $T_e$ than the lattice temperature $T$.
In this model Dirac fermions are assumed to have the usual Fermi-Dirac
distribution $f(E_{\bf k})=[\exp\{\beta_e(E_{\bf k}-\mu)\}+1]^{-1}$ where
$\beta_e=(k_BT_e)^{-1}$ and $\mu$ is the chemical potential determined by the electron density
$n_e=\int f(E_{\bf k})D(E_{\bf k}) dE_{\bf k}$. The 3D Dirac fermions are assumed to interact with the
$3$D phonons of energy $\hbar\omega_{\bf q}$ and wave vector ${\bf q}$. The cooling power per electron
$P$ (i.e. average electron energy loss rate) due to el-ph interaction can be obtained by
using the well known technique described in Ref. [\onlinecite{Conwell}]. It is given by
\begin{eqnarray}\label{cool_rate}
 P=-\frac{1}{N_e}\sum_{\bf q} \hbar\omega_{\bf q} \Bigg(\frac{dN_{\bf q}}{dt}\Bigg)_{\rm el\textendash ph},
\end{eqnarray}
where $N_e$ is total number of electrons,
$N_{\bf q}$ is the non-equilibrium phonon 
distribution function. The rate of change
of $N_{\bf q}$ due to electron-phonon interaction
i.e.$\big(dN_{\bf q}/dt\big)_{\rm el\textendash ph}$ is given by using Fermi's golden rule
\begin{small}
\begin{eqnarray}\label{phRate1}
 \Bigg(\frac{dN_{\bf q}}{dt}\Bigg)_{\rm el\textendash ph}&=&\frac{2\pi \rm g}{\hbar}
 \sum_{\bf k} \vert M({\bf q})\vert^2 \Big\{(N_{\bf q}+1)f(E_{\bf k}+\hbar\omega_{\bf q})\nonumber\\
 &\times&\big[1-f(E_{\bf k})\big]
 -N_{\bf q}f(E_{\bf k})\big[1-f(E_{\bf k}+\hbar\omega_{\bf q})\big]\Big\}\nonumber\\
&\times& \delta(E_{{\bf k}+{\bf q}}-E_{\bf k}-\hbar\omega_{\bf q}),
\end{eqnarray}
\end{small}
where $\vert M({\bf q})\vert^2=\vert g(q)\vert^2\vert F(\theta_{{\bf k}, {\bf k^\prime}})\vert^2$
is the square of the matrix element for the el-ph interaction. Here, $\vert g(q)\vert^2$ is square 
of el-ph matrix element without chiral wave function and 
$\vert F(\theta_{{\bf k}, {\bf k^\prime}})\vert^2=(1+\cos\theta_{{\bf k}{\bf k^\prime}})/2$
with $\theta_{{\bf k}{\bf k^\prime}}$ being the angle between ${\bf k}$ and ${\bf k^\prime}$, 
resulting from the chiral nature of the Dirac fermion.

One may also write Eq.(\ref{phRate1}) in the following form
\begin{eqnarray}\label{phRate2}
 \Bigg(\frac{dN_{\bf q}}{dt}\Bigg)_{\rm el-ph}=\Big[(N_{\bf q}+1)e^{-\beta_e \hbar \omega_{\bf q}}-N_{\bf q}\Big]
 \Gamma_{\bf q},
\end{eqnarray}
where $\Gamma_{\bf q}$ is given by
\begin{eqnarray}\label{Gamma_q}
 \Gamma_{\bf q}&=&\frac{2\pi {\rm g}}{\hbar}\sum_{\bf k} \vert M({\bf q})\vert^2
 f(E_{\bf k})\big[1-f(E_{\bf k}+\hbar\omega_{\bf q})\nonumber\\
 &\times&\delta(E_{{\bf k}+{\bf q}}-E_{\bf k}-\hbar\omega_{\bf q}).
\end{eqnarray}

As a result the cooling power(Eq. (\ref{cool_rate})) becomes
\begin{eqnarray}\label{cool_rate2}
P=\frac{1}{N_e}\sum_{\bf q} \hbar\omega_{\bf q}\Big[(N_{\bf q}+1)e^{-\beta_e \hbar \omega_{\bf q}}-N_{\bf q}\Big]
\Gamma_{\bf q}.
\end{eqnarray}

Our objective is to find hot electron cooling power $P_{\rm op}$ due to optical phonons.
The optical phonon energy $\hbar\omega_{\bf q}=\hbar\omega_0$ is taken to be constant.
The summation over ${\bf q}$ in Eq. (\ref{cool_rate2}) can be converted into integral as 
$\sum_{\bf q}\rightarrow (V/8\pi^3)\int_0^\infty q^2dq\int_0^\pi \sin\varphi d\varphi \int_0^{2\pi}d\psi$ with 
$\varphi$ and $\psi$ are the polar and azimuthal angle of ${\bf q}$, respectively.
Note that the integrations over $\varphi$ and $\psi$ give $4\pi$. Defining $E_q=\hbar v_F q$,
we find hot electron cooling power as
\begin{small}
\begin{eqnarray}
P_{\rm op}=\frac{\hbar\omega_0}{2\pi^2 n_e\hbar^3v_F^3}
\int_0^\infty dE_q E_q^2 \big[(N_q+1)e^{-\beta_e\hbar\omega_0}-N_q\big]\Gamma_q.
\end{eqnarray}
\end{small}
An explicit evaluation of
$\Gamma_q$ is given in the section II(D).

\subsection{Hot phonon distribution}
Non-equilibrium phonon distribution $N_{\bf q}$ can be obtained from the Boltzmann equation
\begin{eqnarray}\label{Boltz}
 \Bigg(\frac{dN_{\bf q}}{dt}\Bigg)_{\rm el\textendash ph}+\Bigg(\frac{dN_{\bf q}}{dt}\Bigg)_{\rm oth}=0,
\end{eqnarray}
where the first term describes the rate of change of the phonon distribution due to 
electron-phonon interaction while the later one denotes the same due to the 
other processes namely phonon-phonon interaction, surface roughness scattering etc.

In the relaxation time approximation one can write 
\begin{eqnarray}
\Bigg(\frac{dN_{\bf q}}{dt}\Bigg)_{\rm oth}=-\frac{N_{\bf q}-N_{\bf q}^0}{\tau_p},
\end{eqnarray}
where, $N_{\bf q}^0=[\exp(\beta \hbar\omega_0)-1]^{-1}$ with $\beta=(k_BT)^{-1}$
is the phonon distribution at equilibrium and $\tau_p$ is 
phonon relaxation time due to all other mechanisms.

Hence, the non-equilibrium phonon distribution $N_{\bf q}$ will be readily obtained from 
Eq.(\ref{Boltz}) as
\begin{eqnarray}\label{Non_q}
N_{\bf q}=N_{\bf q}^0+\tau_p\Bigg(\frac{dN_{\bf q}}{dt}\Bigg)_{\rm el\textendash ph}.
\end{eqnarray}

Inserting Eq. (\ref{phRate2}) into Eq. (\ref{Non_q}) and considering $\hbar\omega_{\bf q}=\hbar\omega_0$,
we find $N_{\bf q}$ as 
\begin{eqnarray}\label{hot_phon}
N_{\bf q}=\frac{N_{\bf q}^0+\tau_p\Gamma_{\bf q}e^{-\beta_e \hbar \omega_0}}
{1+\tau_p\Gamma_{\bf q}(1-e^{-\beta_e \hbar \omega_0})}.
\end{eqnarray}

\subsection{Evaluation of $\Gamma_q$}
Here, we shall provide an explicit evaluation of el-ph scattering
rate $\Gamma_q$. We assume the electron-optical phonon interaction
via Fr\"{o}hlich coupling with the corresponding matrix element 
$\vert g(q)\vert^2=2\pi e^2\hbar\omega_0(\varepsilon_\infty^{-1}
-\varepsilon_0^{-1})/(Vq^2)$, where $\varepsilon_\infty (\varepsilon_0)$
is the high frequency (static) dielectric constant of the material.
Momentum conservation ${\bf k^\prime}={\bf k}+{\bf q}$ also allow us to write  
$\vert F(\theta_{{\bf k}, {\bf k^\prime}})\vert^2$ as
\begin{eqnarray}
 \vert F(\theta_{{\bf k}, {{\bf k}+{\bf q}}})\vert^2
 =\frac{1}{2}\Bigg(1+\frac{E_k+E_q\cos\theta}{E_k+\hbar\omega_0}\Bigg).
\end{eqnarray}

Converting the summation over ${\bf k}$ into integrals like
$\sum_{\bf k}\rightarrow V/((2\pi)^3)\int dk k^2 \sin\theta d\theta d\phi$,
Eq.(\ref{Gamma_q}) becomes
\begin{eqnarray}\label{Gamma3}
\Gamma_{\bf q}&=&\frac{2\pi \rm g}{\hbar}\frac{V}{(2\pi)^3}\int k^2 dk dx d\phi \vert g({q})\vert^2
\vert F(\theta_{{\bf k}, {{\bf k}+{\bf q}}})\vert^2\nonumber\\
&\times& f(E_{\bf k}) \big[1-f(E_{\bf k}+\hbar\omega_0)\big] \delta[X(x)],
\end{eqnarray}
where the argument of the delta function is
\begin{eqnarray}\label{Xe}
 X(x)=(E_k^2+E_q^2+2E_kE_qx)^{\frac{1}{2}}-E_k-\hbar\omega_0
\end{eqnarray}
with $x\equiv\cos\theta$.
Note that the $\phi$-integral in Eq. (\ref{Gamma3}) gives $2\pi$ 
since the integrand is independent of $\phi$. The $x$-integral can be 
evaluated by the following property of the delta function
\begin{eqnarray}
\delta[X(x)]=\frac{\delta(x-x_i)}{\frac{dX}{dx}\big\vert_{x=x_i}},
\end{eqnarray}
where the root $x_i$ of Eq.(\ref{Xe}) can be obtained as
\begin{eqnarray}
x_i=\frac{(\hbar\omega_0)^2+2E_k\hbar\omega_0-E_q^2}{2E_kE_q}. 
\end{eqnarray}

Hence, after doing the angular integrations,
Eq. (\ref{Gamma3}) becomes
\begin{eqnarray}\label{Gamma_4}
\Gamma_{q}&=&{\rm g}\frac{V\vert g({q})\vert^2}{2\pi\hbar(\hbar v_F)^3 E_q}\int_{E_k^{\rm min}}^\infty dE_k E_k 
  \vert F(E_k,E_q)\vert^2\nonumber\\
  &\times& f(E_{\bf k}) \big[1-f(E_{\bf k}+\hbar\omega_0)\big]
(E_k+\hbar\omega_0),
\end{eqnarray}
where $E_k^{\rm min}=(E_q-\hbar\omega_0)/2$. This lower limit of $E_k$-integral is a consequence of 
the fact that $-1\leq x_i\leq1$. Note also that
\begin{eqnarray}
 \vert F(E_k,E_q)\vert^2=\frac{1}{2}\Bigg[1+\frac{(E_k+\hbar\omega_0)^2+E_k^2-E_q^2}{2E_k(E_k+\hbar\omega_0)}\Bigg].
\end{eqnarray}

\subsection{Hot electron cooling power due to acoustic phonons in $3$DDS}
As we are giving results for hot electron cooling power $P_{\rm ac}$ due to acoustic phonons,
for the sake of comparison with $P_{\rm op}$, the corresponding expression is given by\cite{KS_Bhargv}
\begin{eqnarray}
P_{\rm ac}&=&-\frac{{\rm g}D^2}{8\pi^3\rho_m \hbar^7 n_e v_F^4 v_s^4}
\int_0^\infty dE_k \nonumber\\
&\times& \int_0^{\hbar\omega_q^m}d(\hbar\omega_q)(\hbar\omega_q)^3
\frac{(E_k+\hbar\omega_q)^2}{\vert \epsilon(q,T)\vert^2} \chi(q,k)\nonumber\\
&\times& [N_q(T_e)-N_q(T)][f(E_k)-f(E_k+\hbar\omega_q)],
\end{eqnarray}
where $D$ is the acoustic phonon deformation potential coupling constant,
$v_s$ is the acoustic phonon velocity, $\rho_m$ is the mass density,
$\chi(q,k)$=$1-q^2/(4k^2)$, $\omega_q$=$v_sq$, and $\omega_q^m=2v_sk$.
The temperature dependent screening function is given by 
$\epsilon(q,T)=1+\Pi(q,T)$, where $\Pi(q,T)$ is the finite temperature static
polarizability, evaluated explicitly in Ref.[\onlinecite{SD_Sarma}].

\subsection{Hot electron cooling power due to optical phonons in $3$D semiconductor}
With a view to compare the results of the $3$DDS Cd$_3$As$_2$ with those in $3$D Cd$_3$As$_2$
semiconductor, we give the expression for the $P_{\rm op}$ for a 3DEG in bulk Cd$_3$As$_2$ semiconductor. 
Taking the parabolic energy relation $E_k=\hbar^2k^2/(2m^\ast)$ ($m^\ast$ is the effective mass of electron),
it is  given by
\begin{eqnarray}
P_{\rm op}&=&\frac{{m^\ast}^{\frac{3}{2}}\omega_0}{\sqrt{2}\pi^2 n_e \hbar^2}
\int_0^\infty dE_q \,E_q^{\frac{1}{2}}\nonumber\\
&\times&\big[(N_q+1)e^{-\beta_e\hbar\omega_0}-N_q\big]\Gamma_q.
\end{eqnarray}
In this case $E_q=\hbar^2 q^2/(2m^\ast)$. The corresponding el-ph 
scattering rate $\Gamma_q$ is given by
\begin{small}
\begin{eqnarray}
\Gamma_q={\rm g}\frac{V{m^\ast}^{\frac{3}{2}}\vert g(q)\vert^2}{2^{\frac{3}{2}} \pi \hbar^4 E_q^{\frac{1}{2}}}
\int_{E_k^{l}}^\infty dE_k f(E_{\bf k}) \big[1-f(E_{\bf k}+\hbar\omega_0)\big],
\end{eqnarray}
\end{small}
where $E_k^l=(\hbar\omega_0-E_q)^2/(4E_q)$.

\subsection{Electron energy relaxation time}
Some times it is useful to study the hot electron relaxation in terms of corresponding 
relaxation time $\tau_e$,
given by\cite{Ridley_3} $\tau_e=[\la E_k(T_e)\ra-\la E_k(T)\ra]/P$
where $\la E_k(T_e)\ra=(1/N_e)\int E_k f(E_k) D(E_k)dE_k$ is 
the average energy of the electron at temperature $T_e$. For large electron density and very low 
temperature, expressing density of  states $ D(E_k)=D_0 E_k^p$,
we find
\begin{eqnarray}
 \la E_k(T_e)\ra=\frac{p+1}{p+2}E_F\Bigg[1+\frac{(p+2)}{6}\Bigg(\frac{\pi k_B T_e}{E_F}\Bigg)^2\Bigg],
\end{eqnarray}
where $E_F$ is the Fermi energy at $T_e$=$0$. This formula is applicable to $3$D and $2$D fermions 
and Dirac fermions with the respective choice of $E_F$ and $p$.

\section{Results and Discussions}
In the following, we numerically  study the hot phonon distribution and electron cooling power
in $3$D Dirac semimetal Cd$_3$As$_2$ in the range for electron temperature $T_e$=$(5$-$300)$ K and
electron density $n_e$=$(0.1$-$3)n_0$, where $n_0$=$1.0\times10^{24}$ m$^{-3}$, taking the lattice temperature
$T$=$4.2$ K. In addition, we present some results of acoustic phonon limited hot electron 
cooling power in Cd$_3$As$_2$. For comparison we also highlight on the electron cooling
power in conventional 3DEG. The values of the parameters appropriate for numerical calculations are
given in Table I.
\begin{table}[ht]
 \centering
\begin{tabular}{ p{4.5cm} p{1.2cm} p{2.3cm} }
\hline
\hline
Parameter  & Symbol &~~ Value\\
\hline
\hline
Lattice constant &~~ $a$ &~~ 4.6 Angstrom\\
Effective mass of electron\cite{Jay_Gerin} &~~ $m^{\ast}$  &~~ $0.036m_e$\\
Mass density of ion &~~ $\rho$ &~~ $7\times10^3$ Kg/m$^3$\\
Degeneracy    & & \\
~~~For 3DDS&~~ g  &~~ 4\\
~~~For 3DEG&~~ g  &~~ 2\\
Sound velocity &~~ $v_s$  &~~ $2.3\times10^3$ m/s\\
Fermi velocity\cite{Lundgren_2} &~~ $v_F$  &~~ $10^6$ m/s\\
Deformation potential\cite{Jay_Gerin} &~~ $D$ &~~ 20 eV\\
Optical phonon energy\cite{W_Lu, Weszka} &~~ $\hbar\omega_0$&~~ $25$ meV\\
 & &\\
Dielectric constant(High Freq.) &~~ $\varepsilon_\infty$&~~ 12\\
Dielectric constant(Static)   &~~ $\varepsilon_0$&~~ 36\\

Electron density &~~ $n_e$ &~~ $(0.1$-$3)n_0,$\\ 
Lattice temperature &~~ $T$ &~~ $4.2$ K\\

Electron temperature &~~ $T_e$ &~~ $(5$-$300)$K\\
\hline
\hline
\end{tabular}
\caption{Numerical values of the parameters used for calculation. Note that the knowledge of 
the lattice constant $a$ is required to calculate $\Pi(q,T)$, given in Ref. [\onlinecite{SD_Sarma}].}
\end{table}

Hot phonon distribution $N_q$ as a function of $q$ is shown for different $\tau_p$, $n_e$, and $T_e$,
respectively, in Fig. 1 for $3$DDS Cd$_3$As$_2$. In all the three figures we notice that $N_q$ 
varies considerably and this $q$ dependence is determined by $\Gamma_q$ arising
from the  electron-optical phonon interaction in Eq.(\ref{Gamma_4}).
Each curve has a broad maximum for about $q$ = $3.98\times10^7$ m$^{-1}$
where the phonon heating is found to be significant. This
corresponds to the phase matching phonon wave vector $q_0$= $3.79\times10^7$ m$^{-1}$ given by
$\hbar v_F q_0=\hbar\omega_0$. The initial steep increase of $N_q$ may be attributed to the lower
limit of $E_k$ integration in the Eq.(\ref{Gamma_4}). After a broad maximum, with further
increase of $q$, the $N_q$ is found to decrease first very slowly and then gradually
to  zero. Similar observations are made in GaAs heterojunctions\cite{ZJ_Zhang2}, bilayer graphene\cite{Katti}
and monolayer MoS$_2$\cite{Kaasbjrg}. Writing $N_q$ as the hot phonon number $N_q(T_{\rm ph})$ given by Bose distribution at
an effective  hot phonon temperature $T_{\rm ph}$, it can be shown from Eq. (\ref{hot_phon})
that, for $\Gamma_q\gg\tau_p^{-1}$, $T_e\gg T$, and $\beta_e\sim (\hbar\omega_0)^{-1}$ the
$N_q(T_{\rm ph})$ approaches the $N_q(T_e)$.


\begin{figure*}[ht!]
\begin{center}$
\begin{array}{cc}
\includegraphics[height = 50mm,width=140mm]{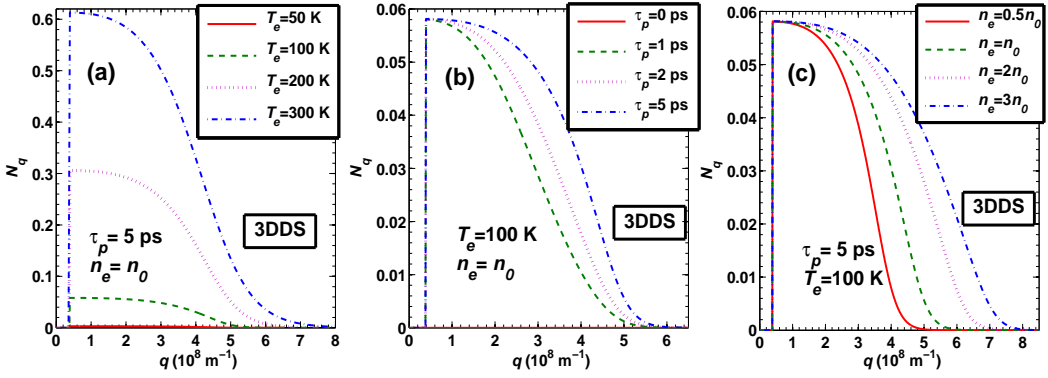}
\end{array}$
\caption{Non-equilibrium distribution of polar optical phonon $N_q$ in 3DDS Cd$_3$As$_2$
as a function function of phonon wave vector $q$ for different $T_e$, $\tau_p$, and $n_e$.}
 \end{center}
\label{Fig2}
\end{figure*}

\begin{figure*}[ht!]
\begin{center}$
\begin{array}{cc}
\includegraphics[height = 50mm,width=140mm]{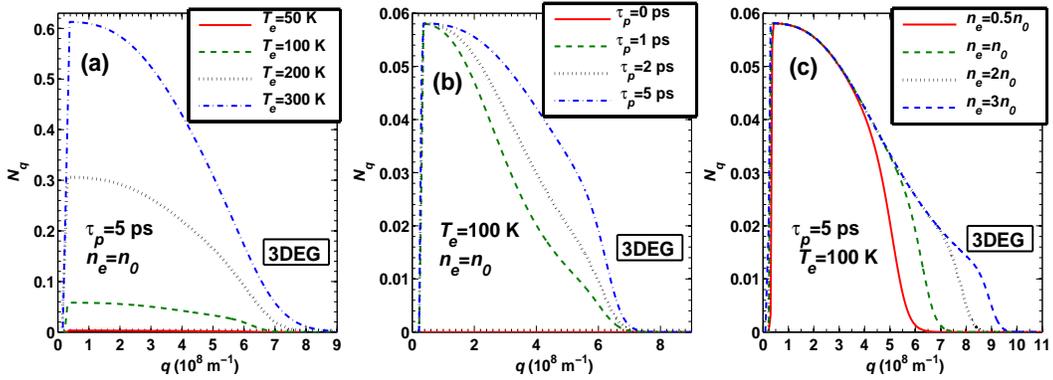}
\end{array}$
\caption{Non-equilibrium distribution of polar optical phonon $N_q$ in 3DEG Cd$_3$As$_2$ semiconductor
as a function function of phonon wave vector $q$ for different $T_e$, $\tau_p$, and $n_e$.}
 \end{center}
\label{Fig2}
\end{figure*}

In Fig. 1(a), $N_q$ vs $q$ is shown for different $T_e$=$50$, $100$, $200$, and $300$ K
with $n_e$=$n_0$ and $\tau_p$=$5$ ps. The phonon number $N_q$ increases with increasing $T_e$,
as expected on the physical ground that electrons with larger $T_e$ can emit large number of phonons.
The $N_q$ corresponding to $T_e$=$50$ K is very small compared to the other $T_e$ values,
with largest value of $N_q\simeq 0.6$ corresponding to $300$K. The width of the maximum is larger
for smaller $T_e$, similar to the findings in monolayer MoS$_2$\cite{Kaasbjrg}.

$N_q$ dependence on phonon relaxation time $\tau_p$ is shown in Fig. 1(b) by plotting
$N_q$ vs $q$ for $\tau_p$= $0$, $1$, $2$, and $5$ ps with $n_e$=$n_0$ and $T_e$=$100$ K. Number of hot phonons
is found to be  larger for larger relaxation time, as anticipated. Secondly, we notice
that the range of $q$ for which $N_q$ remains maximum and nearly constant (i.e. width of the maximum)
is larger for phonons with larger $\tau_p$.

The effect of $n_e$ on hot phonon distribution is shown in Fig. 1(c) for $\tau_p$=$5$ ps at $T_e$=$100$ K.
The electron density chosen are $n_e$=$0.5n_0$, $n_0$, $2n_0$, and $3n_0$. $N_q$ is found to be larger
for larger $n_e$ because larger density of electrons emit large number of phonons. 
Moreover, width of the maximum is larger for larger $n_e$. Similar observation is made in
GaN/AlGaN heterostructure\cite{Gokden}.

For comparison we have shown $N_q$ vs $q$, in Fig. 2 for different $T_e$, $\tau_p$, and 
$n_e$ for $3$DEG in bulk Cd$_3$As$_2$ semiconductor. We see that maximum of $N_q$ in this
semiconductor is almost same as found for $3$DDS Cd$_3$As$_2$ for each of the $T_e$.
But, unlike in $3$DDS Cd$_3$As$_2$, this maximum is spread over relatively a small range of $q$.
The maximum occurs around $q\sim1\times10^{8}$ m$^{-1}$ which is closer to the phase matching value of 
$q_0=(2m\omega_0/\hbar)^{1/2}$= $1.54\times10^8$ m$^{-1}$. In the low $q$ region $N_q$ increases rapidly
with $q$, and after reaching the maximum it gradually decreases. Finally it vanishes at relatively
larger $q$ values as compared to  3DDS.
From Fig. 2(c), in which $N_q$ is shown for different $n_e$, we see that, for Cd$_3$As$_2$
semiconductor $N_q$ remains same, unlike the case of $3$DDS Cd$_3$As$_2$ ( Fig. 1c),
for certain range of $q$ for all $n_e$. Beyond this, $N_q$ is found to be larger for larger $n_e$ at larger $q$.

In Fig. 3(a), we have shown electron cooling power $P_{\rm op}$, due to optical phonons, as
a function of $T_e$ for different phonon relaxation time $\tau_p$=$0$, $1$, $2$, and $5$ ps.
The curve for $\tau_p=0$ ps corresponds to $P_{\rm op}$ without hot phonon effect.
In  low $T_e$ (about $< 50$ K), all the curves show rapid increase
of $P_{\rm op}$ with $T_e$. This is expected because in low
$T_e$ region $\hbar\omega_0/(k_BT_e)$ is large and it decreases significantly with increase of $T_e$.
Consequently the optical phonon emission increases as $\sim \exp(-\hbar\omega_0/(k_BT_e))$.
In the higher $T_e$ region, $P_{\rm op}$ increases slowly.
This behavior may be approximately put as $\exp(-\hbar\omega_0/(k_B T_e))$ attributing to the exponential
growth of occupation of electron states with high enough energy to emit optical phonons.
At $T_e$=$300$ K, the  $\hbar\omega_0/(k_BT_e)$ is nearly
$1$ and $P_{\rm op}$ is nearly constant.
\begin{figure}[h!]
\begin{center}\leavevmode
\includegraphics[width=100mm,height=50mm]{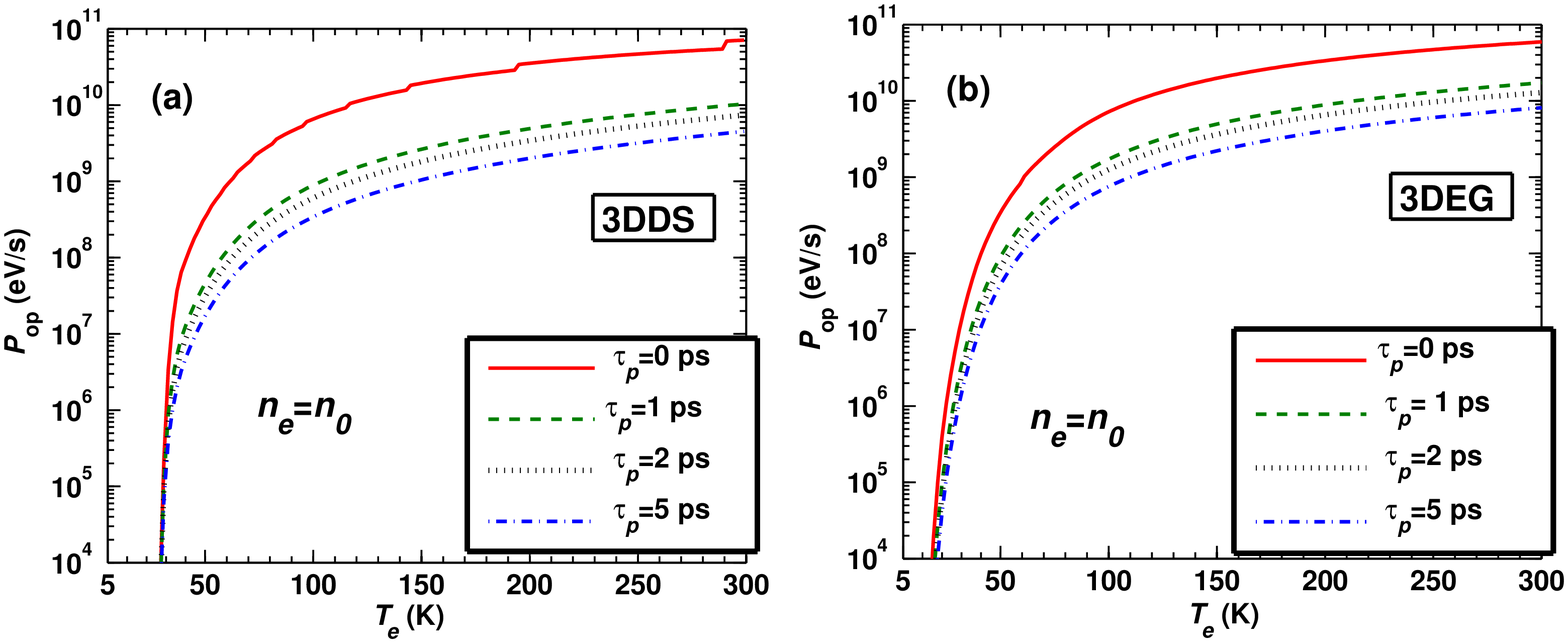}
\caption{Temperature dependence of $P_{\rm op}$ for different values of $\tau_p$ at 
$n_e$=$n_0$. Left and right panels show the results corresponding to 
$3$DDS Cd$_3$As$_2$ and for $3$DEG in Cd$_3$As$_2$ semiconductor, respectively.}
\end{center}
\end{figure}

We find that the hot phonon  effect reduces $P_{\rm op}$ significantly.
However, the hot phonon effect is larger for  $\tau_p<1$ ps and smaller for higher values of $\tau_p$.
We can introduce a reduction factor
$R_{\rm op}$ = $P_{\rm op}$(without hot phonon effect)/$P_{\rm op}$(with  hot phonon effect),
which is always expected to be greater  than $1$. The reduction factor increases with increasing $\tau_p$,
as expected, because there is increased number of hot phonons for larger $\tau_p$ which are partially
reabsorbed and hence reducing the further electron power loss. For example, for $\tau_p$=$1$ ps,
at $T_e$=$100$ and $300$ K, the reduction factors are $R_{\rm op}$=$7.37$ and $6.86$, respectively.
For $\tau_p$=$5$ ps, at $T_e$=$100$ and $300$ K, the reduction factors, respectively,
are $R_{\rm op}$=$19.21$ and $15.72$.
This observation indicates that hot phonon effect is sensitive
at low $T_e$ and less sensitive at higher $T_e$.

For comparison, $P_{\rm op}$ is shown as a function of $T_e$, in Fig. 3(b), for 3DEG in
Cd$_3$As$_2$ semiconductor for $\tau_p$ = $0$, $1$, $2$, and $5$ ps.
It is found that without hot phonon effect (i.e $\tau_p$=$0$), $P_{\rm op}$ in $3$DEG is
much larger than that in $3$DDS in the temperature regime $T_e<30$ K. Above $T_e\sim40$ K,
the corresponding values of $P_{\rm op}$ are almost same. With hot phonon effect ($\tau_p\neq0$), $P_{\rm op}$ 
in $3$DEG is larger than $P_{\rm op}$ in $3$DDS over the entire range of temperature 
considered. The difference between the values of $P_{\rm op}$ in $3$DEG and $3$DDS is huge 
below $T_e\sim30$ K. However, this difference above $T_e\sim 40$ K is small and it is increasing
with $\tau_p$. In Cd$_3$As$_2$ semiconductor, we find for $\tau_p$=$1$ ps at $T_e$=$100(300)$ K,
$R_{\rm op}$=$4.17(3.44)$. For $\tau_p$=$5$ ps, it is obtained $R_{\rm op}$=$9.52(7.29)$ at
$T_e$=$100(300)$ K. These values of $R_{\rm op}$ are smaller than that found in $3$DDS.
This larger reduction of $P_{\rm op}$ in $3$DDS Cd$_3$As$_2$
indicates that, at a given $T_e$, hot phonon population 
is more in this system compared to $3$D Cd$_3$As$_2$ semiconductor. This can be seen from the more
broader maximum of $N_q$ in the former system.
 
In Fig. 4(a)(4(b)) $P_{\rm op}$ is shown as a function of $T_e$ for different $n_e$ in
$3$DDS ($3$DEG) Cd$_3$As$_2$ taking $\tau_p$=5 ps. In $3$DDS Cd$_3$As$_2$ (Fig. 4(a))
$P_{\rm op}$ is found to be smaller for larger $n_e$. For about $T_e<50$ K, $P_{\rm op}$ is
found to be more sensitive to $n_e$ and the dependence becomes weaker at higher $T_e$.
However, the situation is different 
for $3$DEG in Cd$_3$As$_2$ semiconductor. As depicted in Fig. 4(b) the $n_e$ sensitivity of $P_{\rm op}$
is more in high $T_e$ regime than that in low $T_e$ range.
\begin{figure}[h!]
\begin{center}\leavevmode
\includegraphics[width=102mm,height=45mm]{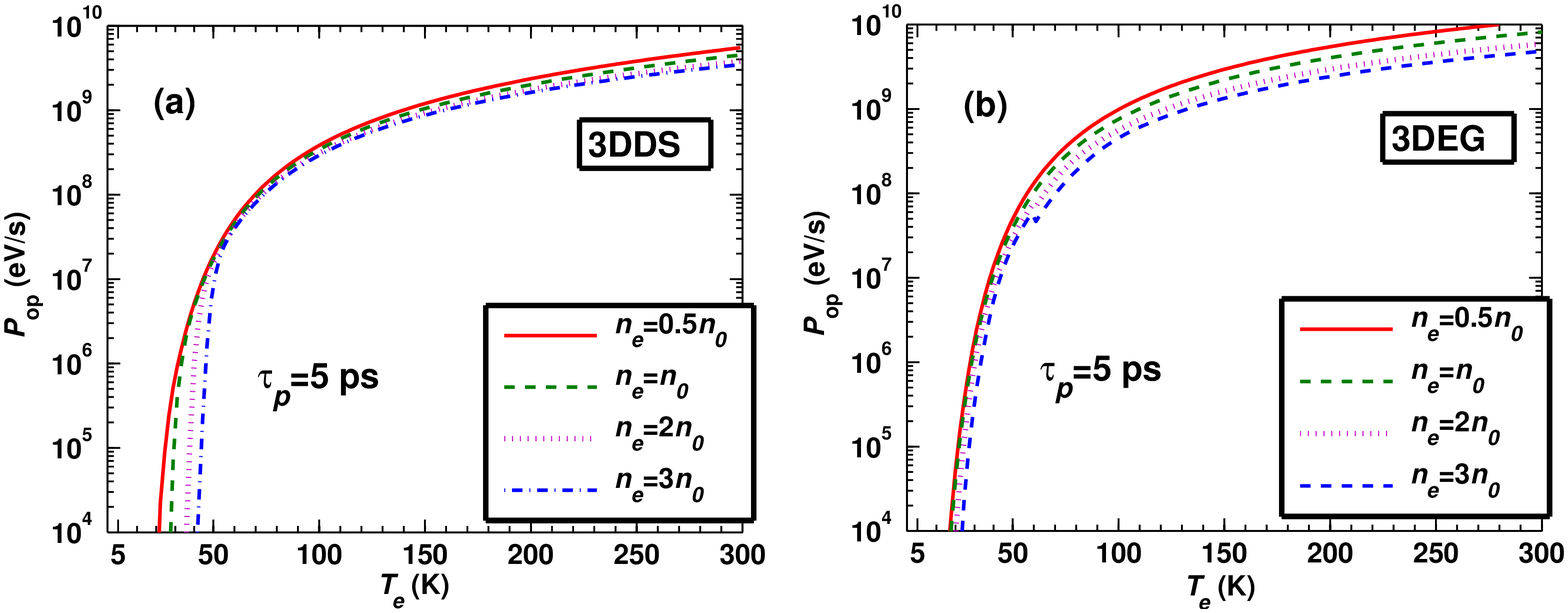}
\caption{Temperature dependence of $P_{\rm op}$ for different values of $n_e$ at $\tau_p$=$5$ ps.
Left and right panels show the results corresponding to 
$3$DDS Cd$_3$As$_2$ and for $3$DEG in Cd$_3$As$_2$ semiconductor, respectively.}
\end{center}
\end{figure}

In Fig. 5, we show temperature dependence of hot electron cooling power due to acoustic
phonons $P_{\rm ac}$ and optical phonons $P_{\rm op}$ considering hot phonon
effect (with $\tau_p$=$5$ ps) as well as the total $P_{\rm T}$=$P_{\rm op}+P_{\rm ac}$.
$P_{\rm ac}$ increases superlinearly in low $T_e$ region and then approaches $\sim T_e$,
which is generic at higher $T_e$. It is also found that, in this temperature range, $P_{\rm ac}$
is smaller for smaller $n_e$, unlike the case of $P_{\rm op}$.
In Fig. 5(b), we see that there is a crossover from acoustic phonon to optical phonon
dominated cooling power. The $T_e$ at which the crossover takes place depends upon $n_e$ significantly.
It is found that crossover takes place at $T_e\sim25$, $30$, $39$, and $45$ K 
for $n_e$=$0.5n_0$, $n_0$, $2n_0$, and $3n_0$, respectively. Optical phonon is the active
channel of power dissipation above this $T_e$. The crossover temperature may depend on $\tau_p$ also.
Considering $n_e$=$0.5n_0$, the crossover $T_e$=$25$ K may be compared with about $20$ K in InSb\cite{InSb}
and $35$ K in GaAs\cite{GaAs,Prabhu} bulk semiconductors noting that these samples are non-degenerate.
The $T_e$=$25$ K above which $P_{\rm op}$ is dominating $P_{\rm T}$ in Cd$_3$As$_2$ is closer to that in bulk InSb as optical
phonon energies in these two systems are closer. It is to be noted that in InSb(GaAs)
$\hbar\omega_0$ is $24.4$($36.5$) meV. In the neighborhood of cross over $T_e$,
$P_{\rm T}$ shows a knee like behavior as found in InSb\cite{InSb} and GaAs\cite{GaAs}.


\begin{figure*}[ht!]
\begin{center}$
\begin{array}{cc}
\includegraphics[height = 50mm,width=140mm]{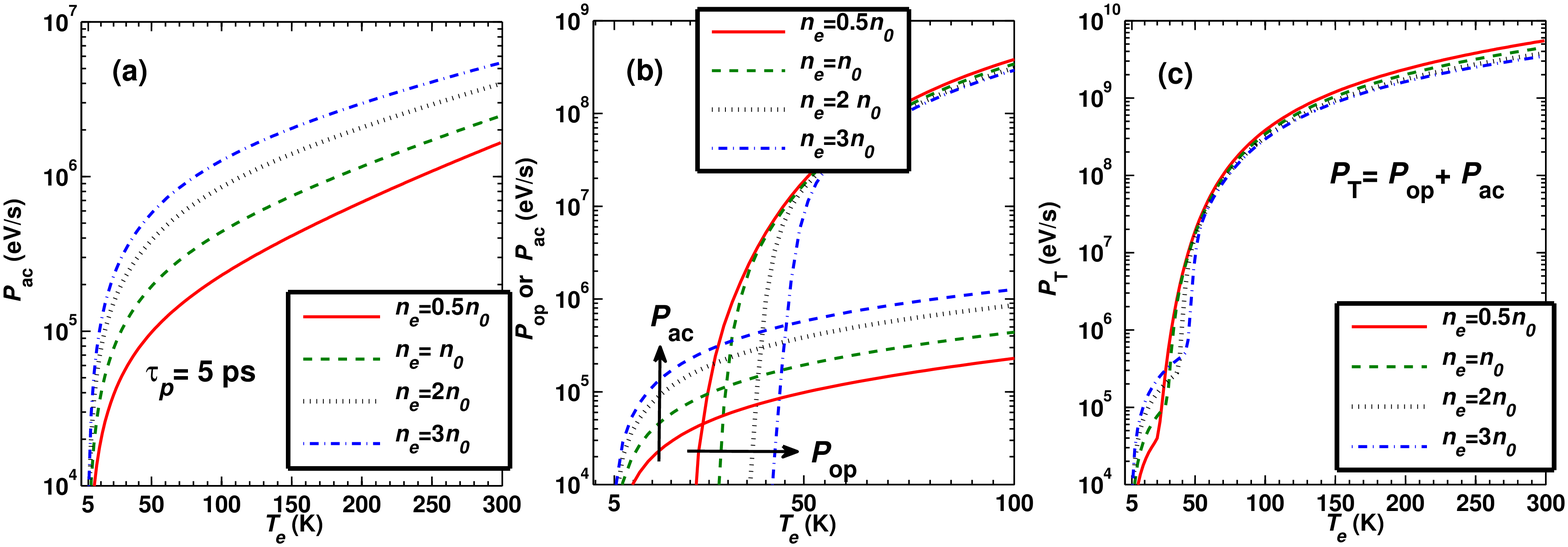}
\end{array}$
\caption{Temperature dependence of $P_{\rm ac}$ and $P_{\rm op}$ in $3$DDS Cd$_3$As$_2$
for different values of $n_e$. (a) $P_{\rm ac}$ in the range $T_e$=$5$-$300$ K,
(b) $P_{\rm ac}$ and $P_{\rm op}$ in the range $T_e$=$5$-$100$ K and
(c) $P_{\rm T}$=$P_{\rm ac}+P_{\rm op}$. Note that $P_{\rm op}$ is for $\tau_p$=$5$ ps.}
 \end{center}
\label{Fig2}
\end{figure*}

In the Bloch-Gr\"{u}neisen (BG) regime, $T_e\ll T_{\rm BG}$=($2\hbar v_s k_F/k_B$),
where $k_F$ is the Fermi wave vector, the $P_{\rm ac}$ dependence on $T_e$ and $n_e$
are shown to be given by the power laws $P_{\rm ac}\sim T_e^\alpha$ and $n_e^{-\delta}$ 
where $\alpha$=$9$($5$) and $\delta$=5/3(1/3) with(without) screening of
electron-acoustic phonon interaction\cite{KS_Bhargv}.
In relatively higher $T_e$ ($>T_{\rm BG}$) regime, disorder assisted $P_{\rm ac}$ calculations
show drastic increase of cooling power due to enhanced energy transfer between electrons and acoustic phonons\cite{Lundgren_2}.

In Fig. 6(a) (6(b)) $P_{\rm op}$ is shown as a function of $n_e$ for different $T_e$
in $3$DDS ($3$DEG) Cd$_3$As$_2$ taking $\tau_p$=$5$ ps. In $3$DDS Cd$_3$As$_2$, $P_{\rm op}$
is found to decrease weakly with increasing $n_e$. This decrease may be attributed to
partial reabsorption of large number of phonons emitted by larger $n_e$.
The decrease is, relatively, faster(slower) at low(high) $n_e$. Moreover, it is found that,
compared to $3$DEG in Cd$_3$As$_2$ semiconductor (Fig. 6(b)), $P_{\rm op}$ in $3$DDS Cd$_3$As$_2$
is less sensitive to $n_e$. This may be attributed to differing density of states.
For example, while varying $n_e$ from $0.1n_0$ to $3n_0$ $P_{\rm op}$ decreases, at $T_e$=$300$ K,
by a factor of about $2.75 (5)$ in $3$DDS ($3$DEG) Cd$_3$As$_2$.
At $T_e$=$100$ K the respective changes are $1.82$ and $4.29$.

\begin{figure}[h!]
\begin{center}\leavevmode
\includegraphics[width=100mm,height=50mm]{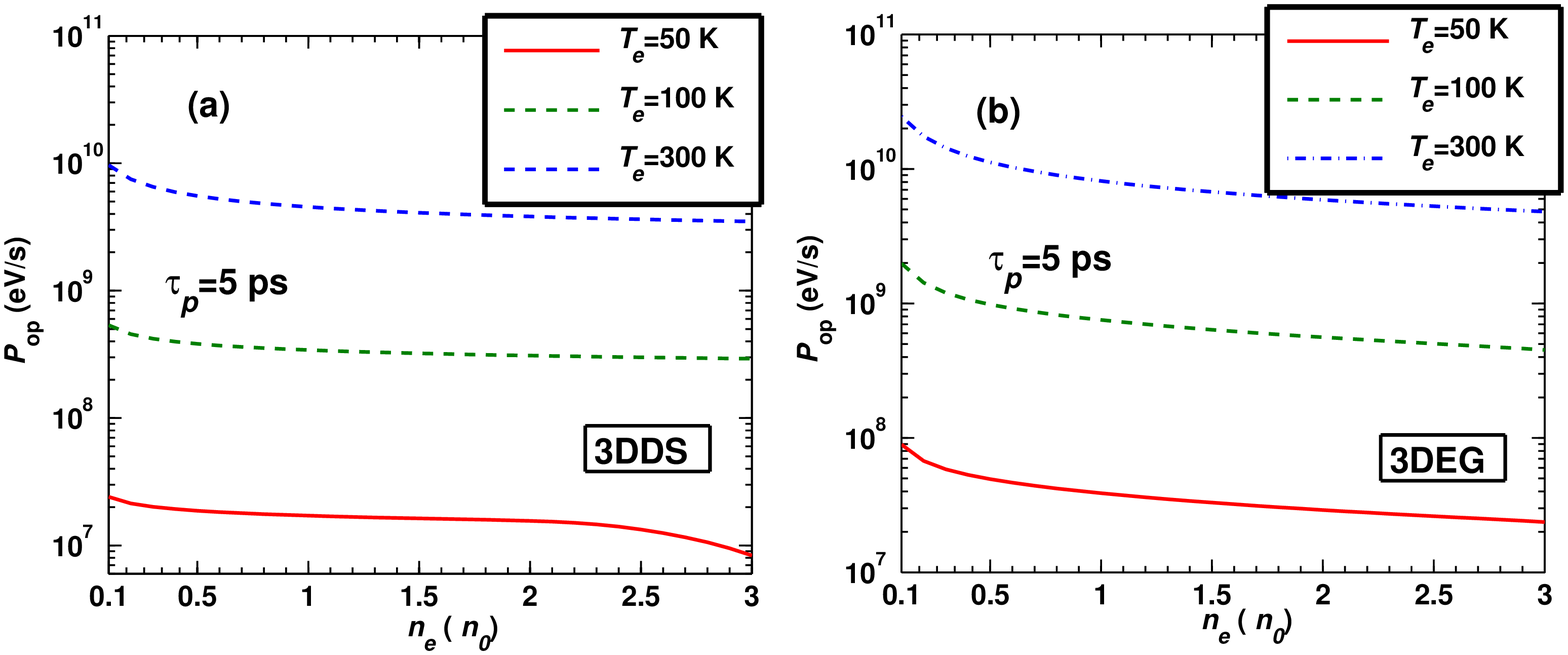}
\caption{Density dependence of $P_{\rm op}$ in $3$DDS Cd$_3$As$_2$ and for $3$DEG 
in Cd$_3$As$_2$ semiconductor for different temperatures.
Hot phonon effect is taken into account with $\tau_p$=$5$ ps.}
\end{center}
\end{figure} 

Electron cooling power due to optical phonons in bulk GaAs is shown to be
reduced by screening effect\cite{SD_Sarma2}.
In $3$DDS Cd$_3$As$_2$ also we expect the screening to reduce $P_{\rm op}$.
Although screening effect is not considered in our $P_{\rm op}$ calculations, 
the experimental measurements will be able to indicate its necessity.

We would like to mention that our numerical calculations of $P$ can be used
to calculate energy relaxation time $\tau_e$ as given in section II(G). For $n_e$=$n_0$, 
the average electron energy is found to be
$\la E(T_e)\ra$=$0.1212$, $0.1235$, and $0.1403$ eV at $T_e$=$4.2$, $100$, and $300$ K,
respectively. Consequently, taking $P_{\rm op}$=$3.417 (45.364)\times 10^8$ eV/s at
$T_e$=$100(300)$ K we find energy relaxation time $\tau_e$=$6.73 (4.21)$ ps.
Alternatively, $\tau_e$ can be obtained from $1/\tau_e$ = $(1/C_e) (dP/dT_e)$, where $C_e$ is the electronic heat capacity.
However, simple power laws can be obtained in BG regime, where $P_{\rm ac}$ is the sole contributor to
electron cooling power, with regard to $T_e$ and $n_e$ dependence. Using the BG regime results,
we find $\tau_e\sim T_e^{-7}$($T_e^{-3}$) and $n_e^{4/3}$($n_e^{0}$) 
for screened (unscreened) electron-acoustic phonon interaction.

\section{Summary}
In summary, we have studied optical phonon limited cooling of hot electrons in $3$DDS Cd$_3$As$_2$
considering the effect of hot phonon. The dependence of electron cooling power $P_{\rm op}$, due to optical 
phonon, on electron temperature $T_e$, electron density $n_e$, and phonon relaxation time $\tau_p$ are 
investigated. $P_{\rm op}$ is found to increase
much rapidly with $T_e$ at low temperature regime while this increase becomes much slower in high $T_e$
regime. The dependence of 
$P_{\rm op}$ on $n_e$  is weak. It shows a slow decrease with the increase of $n_e$.
We compare the results with those corresponding to $3$DEG in Cd$_3$As$_2$ semiconductor.
It is revealed that hot phonon effect is stronger in $3$DDS Cd$_3$As$_2$ than in Cd$_3$As$_2$ semiconductor.
It is also found that $P_{\rm op}$ is more (less) sensitive to $n_e$ in $3$DEG ($3$DDS).
Additionally, $P_{\rm op}$ is compared with the acoustic phonon limited hot electron cooling power $P_{\rm ac}$.
A crossover from $P_{\rm ac}$ dominated cooling at low $T_e$ to $P_{\rm op}$ dominated
cooling at higher $T_e$ takes place at about $T_e$=$25$ K for $n_e$=$0.5n_0$. 
The crossover $T_e$ shifts towards higher temperature for larger $n_e$.
We point out that our calculations need to be tested against the experimental data.
We suggest for steady state/electric field experiments in $n$-type $3$DDS Cd$_3$As$_2$ 
to which our present calculations will be directly related.

\end{document}